\documentclass[twocolumn,preprintnumbers,amsmath,superscriptaddress,amssymb,aps]{revtex4-1}

\usepackage{graphicx}
\usepackage{dcolumn}
\usepackage{bm}
\usepackage{amsmath}
\usepackage{amssymb}
\usepackage{graphicx}
\usepackage{indentfirst}
\usepackage{booktabs}
\usepackage{multirow}
\usepackage{colortbl}

\selectfont

\begin{document}
\title{Stacking, Strain-Engineering Induced Altermagnetism, Multipiezo Effect, and Topological State in Two-Dimensional Materials}
\author{Wei Xun}
\address{Jiangsu Engineering Research Center for Design and Application of Intelligent Detection Equipment, School of Electronic Engineering, Jiangsu Vocational College of Electronics and Information, Huaian 223003, China}
\author{Xin Liu}
\address{Jiangsu Engineering Research Center for Design and Application of Intelligent Detection Equipment, School of Electronic Engineering, Jiangsu Vocational College of Electronics and Information, Huaian 223003, China}
\author{Youdong Zhang}
\email{z.yd@163.com}
\address{Jiangsu Engineering Research Center for Design and Application of Intelligent Detection Equipment, School of Electronic Engineering, Jiangsu Vocational College of Electronics and Information, Huaian 223003, China}
\author{Yin-Zhong Wu}
\address{School of Physical Science and Technology, Suzhou University of Science and Technology, Suzhou 215009, China }
\author{Ping Li}
\email{pli@xjtu.edu.cn}
\address{State Key Laboratory for Mechanical Behavior of Materials, Center for Spintronics and Quantum System, School of Materials Science and Engineering, Xi'an Jiaotong University, Xi'an, Shaanxi, 710049, China}
\address{State Key Laboratory for Surface Physics and Department of Physics, Fudan University, Shanghai, 200433, China}
\address{State Key Laboratory of Silicon and Advanced Semiconductor Materials, Zhejiang University, Hangzhou, 310027, China}

\date{\today}

\begin{abstract}
Altermagnetism, as a newly identified form of unconventional antiferromagnetism, enables the removal of spin degeneracy in the absence of net magnetization that provides a platform for the low power consumption and ultra-fast device applications. However, the rare attention has been paid to the relationship between stacking, strain and altermagnet, multipiezo effect and topological state. Here, we propose a mechanism to realize the altermagnet, multipiezo effect, and topological state in two-dimensional materials by the stacking and strain engineering. Based on the analysis of symmetry, we find that the spin splitting feature related to the $\emph{Ut}$, $\emph{PTt}$, $\emph{M$_z$Ut}$, or $\emph{M$_z$PTt}$ symmetries in altermagnet multilayers. In addition, we find that the stacking engineering can effectively realize the transform from antiferromagnetism to altermagnetism and semiconductor to metal for the Jauns bilayer V$_2$SeTeO. More interestingly, the strain not only induces an intriguing multipiezo effect, encompassing the piezovalley, piezomagnetism and piezoelectric, but also achieves the abundant topological phase. Our findings offer a generalized direction for manipulating the spin splitting, valley polarization, and topological states, promoting practical application of valleytronic and spintronic devices based on two-dimensional altermagnets.
\end{abstract}

\maketitle
\section{Introduction}
Altermagnetism, a new type of collinear magnetism distinguished from antiferromagnetism (AFM) and ferromagnetism (FM), has garnered growing research interest in condensed matter physics \cite{1,2,3,4,5,6}. It exhibits spin splitting similar to that of a ferromagnet, while it forms the AFM-like order with zero net magnetization. The spin splitting in altermagnets arises from the magnetic space group and is safeguarded by crystal symmetry \cite{1,2}. Moreover, the spin dependent Fermi surface of altermagnet shows the planar or bulk $\emph{d}$-wave, $\emph{g}$-wave, or $\emph{i}$-wave symmetry. These characteristics of altermagnetic materials can lead to a variety of novel physical properties. Such as, the anomalous Hall effect with a strength comparable to that of FM \cite{7}; the staggered spin-momentum interaction \cite{8}; the giant and tunneling magnetoresistance effect \cite{9}; the theoretically predicted spin-splitter torque \cite{10} and experimentally proved \cite{11,12}; the piezomagnetism and C-paired spin momentum locking \cite{13}; the nontrivial superconductivity \cite{14,15}; and the anti-Kramers nodal surfaces and unconventional magnetism \cite{16}. Until now, it mainly focuses on the investigated novel physical properties of three-dimensional materials such as RuO$_2$ \cite{5,7,12}, MnTe \cite{4,17,18}, FeSb$_2$ \cite{16}, CrSb \cite{19,20,21}. However, two-dimensional (2D) altermagnetism is rarely reported \cite{22,23,24}, which is in urgent need of systematic investigation.

The layer degree of freedom offer a unique platform for enriching and tuning novel physical phenomena \cite{25,26,27}. For example, the magic-angle graphene superlattices induces unconventional superconductivity \cite{28}. Moreover, the interlayer coupling can tune the magnetic ground state \cite{26,29}. Besides, the layer Hall effect can be realized by layer-dependent engineer in multiferroic lattice \cite{30}. Furthermore, the topological phase transitions can be achieved by changing the interlayer coupling \cite{31}. Therefore, what novel properties are induced by the layer degree of freedom in the altermagnets, that are worth investigating. In addition, strain engineering also provides an opportunity to investigate physics \cite{32,33,34,35}. Under mechanical strain, the 2D materials show intriguing responses including topological phase transition \cite{32,33,34,36}, piezoelectricity \cite{37}, and piezomagnetism \cite{38,39}. The topological phase transition is tuned by strain that is the most common approach \cite{32,33,34,36}. Piezoelectricity, a well-known electromechanical coupling phenomenon, enables the generation of voltage in response to mechanical strain \cite{37}. Besides, piezomagnetism is only observed in certain AFM crystals, where external strain can induce a net magnetization \cite{38,39}. It's worth noting that the valley has been proposed as the third degree of freedom beyond the electron's charge and spin \cite{40,41,42}. Whether the strain can achieve piezovalley phenomenon?

In this work, we propose a mechanism to altermagnet, multipiezo effect, and topological state in 2D materials by the stacking and strain engineering. Firstly, we prove the spin-layer coupling effect in bilayer V$_2$SeTeO. Then, we find that the stacking configuration not only realizes the transform from AFM to altermagnet, but also accompanies the transition from semiconductor to metal. Moreover, a novel multipiezo effect, encompassing the piezovalley, piezomagnetism and piezoelectric, can be achieved in bilayer V$_2$SeTeO. Noted that these two responses are independent of each other, which is different from magnetoelectric coupling. Finally, the Se-Te interface bilayer V$_2$SeTeO exhibits nodal loop under the 0$\%$ $\sim$ -3$\%$ uniaxial strain and the 0$\%$ $\sim$ -2$\%$ biaxial strain, while forms the Weyl points at near X and Y points under the further increased compressive strain. Besides, the Se-Se/Te-Te interface bilayer V$_2$SeTeO becomes Dirac semimetal under the uniaxial/biaxial compressive strain of less than 2$\%$. Our discovery of enriched physical properties in bilayer V$_2$SeTeO offers a platform for designing advanced multifunctional valleytronic and spintronic devices.

\section{STRUCTURES AND COMPUTATIONAL METHODS}
The structure optimization, magnetic, and electronic properties are employed the Vienna $Ab$ $initio$ Simulation Package (VASP) based on the framework of the density functional theory (DFT) \cite{43,44,45}. The generalized gradient approximation (GGA) with the Perdew-Burke-Ernzerhof (PBE) is implemented to describe the exchange-correlation energy \cite{46}. The plane-wave basis with a kinetic energy cutoff is set to be 600 eV. The $21\times 21\times 1$ $\Gamma$-centered $k$ meshes of Brillouin zone is used. A vacuum of 30 $\rm \AA$ is added along the c-axis, to avoid the interaction between the sheet and its periodic images. The total energy and force convergence criterion are set to be -0.005 eV/$\rm \AA$ and 10$^{-8}$ eV, respectively. To describe strongly correlated 3d electrons of V \cite{47}, the GGA + U method is applied with the Coulomb repulsion U value of 4.0 eV. The zero damping DFT-D3 method of Grimme is considered for van der Waals (vdW) correction in bilayer V$_2$SeTeO \cite{38}. The maximally localized Wannier functions (MLWFs) are performed to construct an effective tight-binding Hamiltonian to investigate the topological properties by Wannier90 and WannierTools software package \cite{48,49,50}.

\section{RESULTS AND DISCUSSION }	
\subsection{Structure and symmetry}
As shown in Fig. 1(a), it exhibits the crystal structure of monolayer Janus V$_2$SeTeO, which is a sandwich structure with three atomic layers. The plane formed by V and O atoms is sandwiched by Se and Te planes. The monolayer V$_2$SeTeO shows a tetragonal lattice structure with the space group of P4mm and point group of C$_{4v}$. The calculated lattice constant is 3.93 $\rm \AA$ for monolayer V$_2$SeTeO, while the bond length of V-Se and V-Te is 2.58 $\rm \AA$ and 2.79 $\rm \AA$, respectively. We determine the magnetic ground state of monolayer V$_2$SeTeO by comparing the total energies of two typical magnetic configurations, including FM and AFM states. The AFM configuration energy is 394.39 meV lower than FM configuration, which indicates that the AFM state is the magnetic ground state. However, from the spin charge density of Fig. 1(e), it is not a normal AFM. The two V atoms, which possess opposite spin orientations, are connected by the M$_{xy}$ mirror symmetry. This type of AFM phase is named as altermagnetism.

\begin{figure}[htb]
\begin{center}
\includegraphics[angle=0,width=1.0\linewidth]{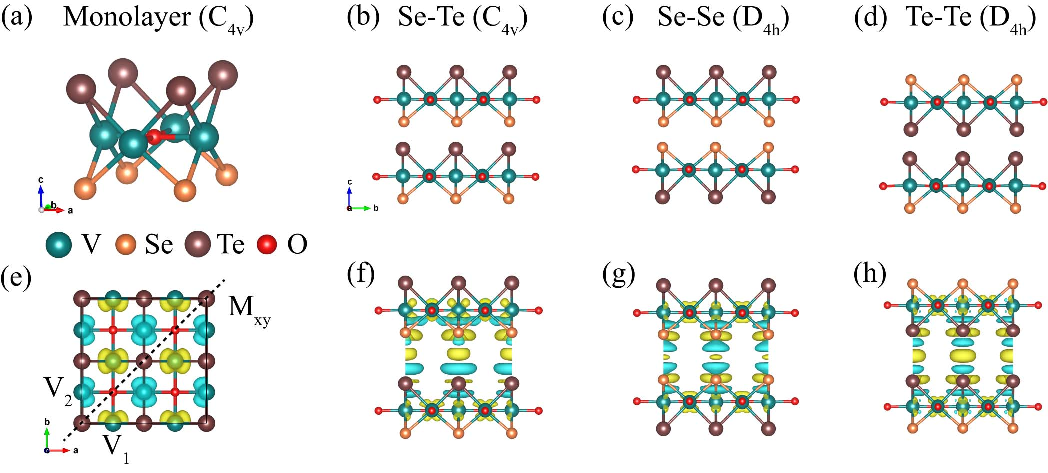}
\caption{(a) The side view of the crystal structure for monolayer V$_2$SeTeO. (b-d) The side view of bilayer V$_2$SeTeO, (b) Se-Te interface, (c) Se-Se interface, and (d) Te-Te interface. (e) Spin charge density of monolayer V$_2$SeTeO. (f-h) The differential charge density of bilayer V$_2$SeTeO, (f) Se-Te interface, (g) Se-Se interface, and (h) Te-Te interface. The yellow and light blue areas denote gain and loss electrons, respectively.
}
\end{center}
\end{figure}

Here, as shown in Fig. 1(b-d), we consider three stacking orders bilayer V$_2$SeTeO, including the upper layer shifted the lower layer formed Se-Te interface bilayer V$_2$SeTeO; the upper layer reversed by the M$_z$ mirror symmetry operation regarding the lower layer formed Se-Se and Te-Te interface bilayer V$_2$SeTeO. It is worth noting that the Se-Se and Te-Te interface bilayer V$_2$SeTeO have the space inversion symmetry ($\emph{P}$), while breaks the $\emph{P}$ symmetry for Se-Te interface bilayer V$_2$SeTeO. Moreover, as shown in Fig. 1(f-h), it can be clearly observed that the Se-Te forms an asymmetric charge transfer, while the strictly symmetric charge transfer is formed in Se-Se and Te-Te interfaces. It indicates that the Se-Te interface bilayer V$_2$SeTeO breaks the combined of time-reversal ($\emph{T}$) and $\emph{P}$ symmetry ($\emph{PT}$), while the Se-Se and Te-Te interface bilayer V$_2$SeTeO possess the $\emph{PT}$ symmetry.

In the AFM system, the energy eigenvalues E$_{\uparrow}$($\textbf{k}$) and E$_{\downarrow}$($\textbf{k}$) are connected, leading to the formation of fully spin-compensated bands. This phenomenon arises from the $\emph{PT}$ symmetry enabled by the two sublattices with opposite spins \cite{51}. The $\emph{P}$ operation only reverses the vector $\emph{k}$ to generate $\emph{P}$E$_{\uparrow}$($\emph{k}$) = E$_{\uparrow}$($\emph{-k}$), and the $\emph{T}$ operation reverses both $\emph{k}$ and spin to generate $\emph{T}$E$_{\uparrow}$($\emph{k}$) = E$_{\downarrow}$($\emph{-k}$). Therefore, the $\emph{PT}$ symmetry guarantees that E$_{\uparrow}$($\emph{k}$) = $\emph{PT}$E$_{\uparrow}$($\emph{k}$) = E$_{\downarrow}$($\emph{k}$), leading to spin degenerate bands for the two spin components with opposite orientations in $\emph{k}$ space. Besides, the translation operation ($\emph{t}$) satisfies $\emph{t}$E$_{\uparrow}$($\emph{k}$) = E$_{\uparrow}$($\emph{k}$). Correspondingly, the energy eigenvalue exists $\emph{PTt}$E$_{\uparrow}$($\emph{k}$) = E$_{\uparrow}$($\emph{k}$) in conventional AFM system. When the spin-orbit coupling (SOC) is ignored, the real space and spin space are completely decoupled. It will lead to $\emph{U}$E$_{\uparrow}$($\emph{k}$) = E$_{\uparrow}$($\emph{k}$), where the $\emph{U}$ is spin reversal operation. Noted that the spin reversal operation $\emph{U}$ is only suitable for collinear systems.

From the analysis discussed above, the existence of $\emph{Ut}$ or $\emph{PTt}$ symmetries in an altermagnet monolayer ensures the spin degeneracy. Due to the lack of an out-of-plane wave vector $\emph{k}$ in low-dimensional system, the energy eigenvalues stay unchanged under the planar mirror symmetry M$_z$, satisfying M$_z$E$_{\uparrow}$($\emph{k}$) = E$_{\uparrow}$($\emph{k}$). Therefore, when opposite spin sublattices are linked by $\emph{Ut}$, $\emph{PTt}$, $\emph{M$_z$Ut}$, or $\emph{M$_z$PTt}$ symmetries, the spin eigenvalues are completely degenerate at any wave vector $\emph{k}$ in low-dimensional system \cite{2,52,53,54,55}. In addition, the diagonal mirror symmetry M$_{\Phi}$ protects the valley degeneracy in the tetragonal structure, indicating that significant valley polarization can be realized through uniaxial strain, which breaks the lattice symmetry.

\subsection{Stacking and strain induced altermagnetism, valley polarization and piezomagnetism}
We propose that layertronics engineered by breaking specific symmetries can effectively control layer-spin locking and valley-contrasting properties, a concept that can be broadly applied to various 2D altermagnetic bilayers. Here, we investigate three typical Janus V$_2$SeTeO stacking structures. The Se-Se and Te-Te interface bilayer V$_2$SeTeO possesses M$_z$ and $\emph{P}$ symmetries, while the Se-Te interface bilayer V$_2$SeTeO breaks M$_z$ and $\emph{P}$ symmetries. Therefore, the $\emph{PTt}$ and $\emph{M$_z$Ut}$ symmetries are broken for the Se-Te interface bilayer V$_2$SeTeO. As shown in Fig. 2(a), the spin degeneracy disappears, showing the characteristics of altermagnetic band structure. Simultaneously, the spin up band of upper layer and spin down band of lower layer cross at the Fermi level X point, while spin down band of upper layer and spin up band of lower layer cross at the Fermi level Y point (see Fig. S1). However, the sublattices of Se-Se and Te-Te interface bilayer V$_2$SeTeO are coupled by the $\emph{PTt}$ and $\emph{M$_z$Ut}$ symmetries. Hence, as shown in Fig. 2(b, c), the band structures exhibit spin degeneracy. Meanwhile, as shown in Fig. S2 and Fig. S3, layer-resolved band structures is also degenerate.

\begin{figure}[htb]
\begin{center}
\includegraphics[angle=0,width=1.0\linewidth]{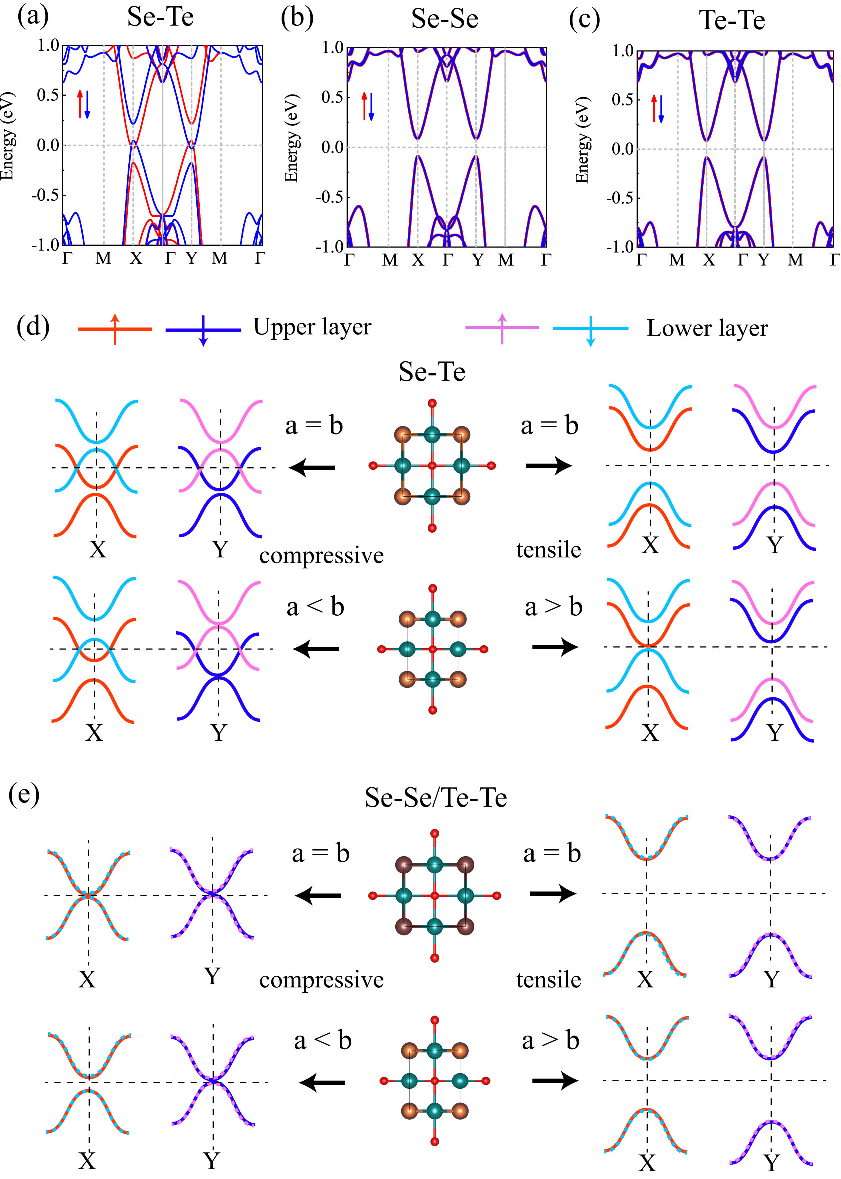}
\caption{(a-c) Spin-polarized band structure of (a) Se-Te interface, (b) Se-Se interface, and (c) Te-Te interface bilayer V$_2$SeTeO. The solid red and blue lines denote spin up and spin down bands, respectively. (d, e) Schematic diagram of the piezovalley mechanism for (d) Se-Te interface, and (e) Se-Se or Te-Te interface bilayer V$_2$SeTeO. The red, blue, magneta, and light blue lines represent spin up, spin down of upper layer and spin up, spin down of lower layer, respectively.
}
\end{center}
\end{figure}

In general, the valley polarization exists in the broken $\emph{P}$ symmetry with ferroelectricity \cite{56} or broken $\emph{T}$ symmetry with the SOC in magnetic system \cite{57,58}. However, the presence of valley polarization in bilayer V$_2$SeTeO relies on uniaxial strain to break the M$_{xy}$ symmetry. This valley polarization can be understood as a response to strain and is named piezovalley. The physical mechanism of piezovalley is shown in Fig. 2(d, e). The valley polarization of bilayer V$_2$SeTeO is defined as the energy difference $\Delta V$ and $\Delta C$ between X and Y at the valence band maximum (VBM) and conduction band minimum (CBM), $\Delta C(V)$ = E$_{Xc(v)}$ - E$_{Yc(v)}$. Fig. 3(a) and Fig. S4(a) exhibit the evolution of valley polarizations and band gaps in Se-Te interface bilayer V$_2$SeTeO under varying uniaxial strain. When the uniaxial strain is less than 2$\%$, the Se-Te interface bilayer V$_2$SeTeO is in a metallic state. Continuing employing the uniaxial tensile strain to 2$\%$, the Y point open a gap, while X point remains closed (see Fig. S5). We named it half-valley metal. At the 5$\%$ uniaxial tensile strain, the valley splitting of VBM and CBM is up to -8.8 meV and 191.8 meV, respectively. In addition, it appears metallic under the biaxial strain of less than 1$\%$, while it exhibits semiconductor at biaxial strains greater than 2$\%$. As shown in Fig. S5, they all exhibit good altermagnetic signatures. These findings demonstrate that the Se-Te interface bilayer V$_2$SeTeO maintains an altermagnetic structure and achieves significantly enhanced piezovalley properties under strain.

\begin{figure}[htb]
\begin{center}
\includegraphics[angle=0,width=1.0\linewidth]{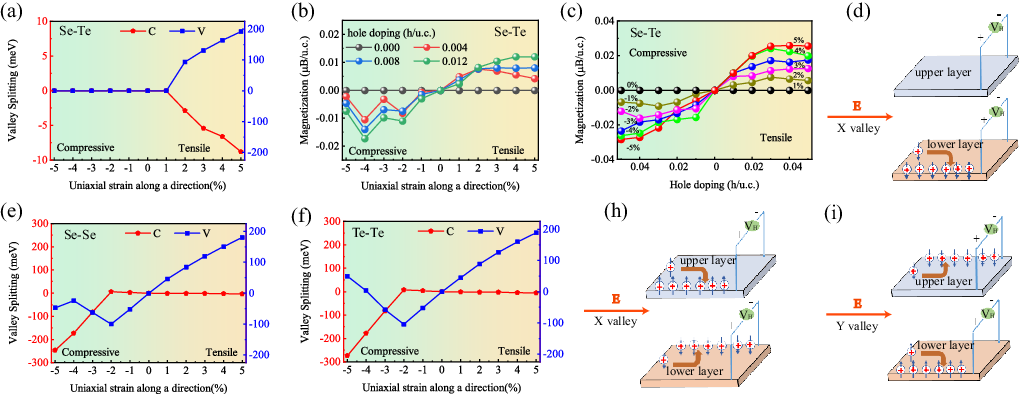}
\caption{(a) Valley splitting of Se-Te interface bilayer V$_2$SeTeO as a function of uniaxial strain along a direction. The valence and conduction bands are labeled V and C, respectively. (b, c) The net magnetization per unit cell as a function of hole doping concentration. (d) Schematic diagram of spin-layer locked anomalous valley Hall effect in the hole-doped Se-Te interface bilayer V$_2$SeTeO at the X valley. The holes are denoted by the + symbol. Downward arrows shows the spin down carriers. (e, f) Valley splitting of (e) Se-Se interface and (f) Te-Te interface bilayer V$_2$SeTeO as a function of uniaxial strain along a direction. The valence and conduction bands are labeled V and C, respectively. (h, i) Schematic diagram of valley layer-spin Hall effect for (h) X valley and (i) Y valley. The holes are shown by the + symbol. Upward and downward arrows represent the spin up and spin down carriers, respectively.
}
\end{center}
\end{figure}

Strain-induced valley splitting in the Se-Te interface bilayer V$_2$SeTeO provides the new approaches to generate net magnetization. The magnetism is determined by integrating the spin density within the energy range from negative infinity up to the Fermi level. The carrier doping can be effectively tuned the Fermi level cross only one valley, leading to the creation of net magnetic moments. The net magnetic moment is defined as M = $\int ^{E_f(n)}_{-\infty}[\rho^{\uparrow}(\varepsilon) - \rho^{\downarrow}(\varepsilon)]$dE, where n, E$_f$, $\rho^{\uparrow(\downarrow)}$, and $\varepsilon$ are the doping density, the doped Fermi level, spin up (spin down) part of the density of states, and external strain, respectively. In the absence of doping, the no net magnetization emerges in the strained Se-Te interface bilayer V$_2$SeTeO, which are consistent with the undoped band structures (see Fig. S5). At the certain doping concentration, the magnetization increases with uniaxial tensile strain, while it oscillates for uniaxial compressive strain. It's worth noting that the induced magnetization is opposite by uniaxial tensile and compressive strains. As shown in Fig. 3(b, c), under the small strain, the magnetization displays a linear response, which gradually saturates as the strain increases. The net magnetic moments generated by the piezomagnetic effect in the Se-Te interface bilayer V$_2$SeTeO are of the same order of magnitude as those previously reported in the monolayer V$_2$Se$_2$O and V$_2$SeTeO \cite{13,38}. The characteristic altermagnetic structure of Se-Te interface bilayer V$_2$SeTeO, coupled with their low magnetocrystalline anisotropy, enable efficient manipulation of magnetic orientation and moments using the electric field and doping. With the hole-doping case, when the Fermi level is shifted between the X and Y valleys of the VBM, the spin down holes of X valley will be generated and accumulated on the left edge of the lower layer [see Fig. 3(d)]. Accordingly, the spin-layer locked anomalous valley Hall effect can be realized.

For the Se-Se and Te-Te interface bilayer V$_2$SeTeO, they exhibit the same pattern, since they all have $\emph{PT}$ symmetry. As shown in Fig. S7 and Fig. S9, the uniaxial compressive strain decreases the band gap at Y point, while the uniaxial tensile strain increases the band gap at Y point. When the uniaxial is less than -2$\%$, both Se-Se and Te-Te interfaces become half-valley-metal. More interestingly, as shown in Fig. 3(e, f), the uniaxial strain can tune not only the magnitude but also the direction of valley polarization. It's worth noting that the uniaxial tensile strain can tune the VBM valley splitting to $\sim$ 200 meV, while the uniaxial compressive strain can regulate CBM valley splitting to $\sim$ -300 meV. This has not been achieved in previous reports. In addition, the band gap of X point, Y point, and total variation are shown in Fig. S4(b, c). On the contrary, the biaxial strain cannot realize valley splitting, but it can achieve semiconductor to metal transition (see Fig. S8 and Fig. S10). For the Se-Se and Te-Te interface in the uniaxial tensile strain, as shown in Fig. 3(h), the spin up and spin down holes of X valley will generated and accumulate on the left edge of upper layer and the right edge of lower layer, respectively. When change from the uniaxial tensile strain to the uniaxial compressive strain, as shown in Fig. 3(i), the hole in the Y-valley will produce the exact opposite. We name the physical phenomenon the valley layer-spin Hall effect.

\subsection{Strain induced topological properties}
The strain realizes rich topological phase transitions. For the Se-Te interface bilayer V$_2$SeTeO, when the uniaxial compressive strain in the 0 $\sim$ -3 $\%$ and the biaxial compressive strain in the 0 $\sim$ -2 $\%$, the band structure form nodal loop near the Fermi level at both the X and Y points, which it consists of different spin channels (see Fig. S5 and Fig. S6). Here, we exhibit the nodal loop of Se-Te interface bilayer V$_2$SeTeO under -1$\%$ biaxial strain in Fig. 4(a). Symmetry analysis uncovers that nodal loops near the X and Y points is protected by the M$_z$ mirror symmetry. The nodal loops are situated in the xy plane, which remains invariant under the M$_z$ mirror symmetry. When the compressive strain is further increased, the Weyl points are formed near X and Y points (see Fig. S5 and Fig. S6). For the Se-Se and Te-Te interface bilayer V$_2$SeTeO, when the uniaxial compressive strain surpasses -2$\%$, the band gap at Y point is closed, while there is still a gap at X point, resulting in the formation a Dirac point. It should be noted that the Dirac points are formed at both X and Y points, when the biaxial compressive strain greater than -2$\%$. Moreover, we calculated the edge state for the Te-Te interface bilayer V$_2$SeTeO. As shown in Fig. 4(b), one can clearly see an edge state connecting two Dirac points. It is worth noting that the Se-Te interface and Se-Se/Te-Te interface bilayer V$_2$SeTeO exhibit the different topological properties due to the unique crystal symmetries in these materials.

\begin{figure}[htb]
\begin{center}
\includegraphics[angle=0,width=1.0\linewidth]{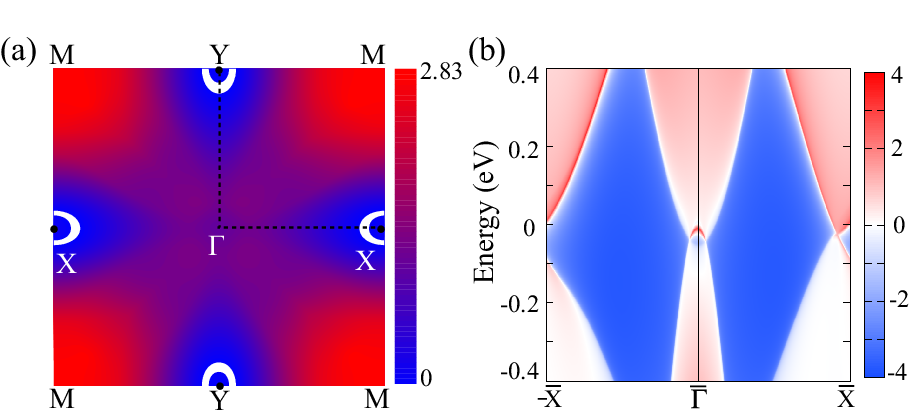}
\caption{ (a) The nodal loop of Se-Te interface bialyer V$_2$SeTeO under -1$\%$ biaxial strain. The color map exhibits the local band gao at the cross of the two bands. (b) The edge state of Te-Te interface bialyer V$_2$SeTeO under -2$\%$ biaxial strain.
}
\end{center}
\end{figure}

\subsection{Stacking and strain induced piezoelectric effect}
The piezoelectric effect, an intrinsic electromechanical coupling phenomenon in noncentrosymmetric materials, arises from strain or stress induces charge redistribution. It can result in the generation of electricity and formation of electric dipole moments. Therefore, as shown in Fig. 5, the Se-Te interface bilayer V$_2$SeTeO shows out-of-plane pieoelectricity due to the broken $\emph{P}$ and M$_z$ symmetries. The second-rank piezoelectric stress tensor e$_{ij}$ and strain tensor d$_{ij}$ can be employed to describe the piezoelectric effect for 2D material. They can be obtained as the total of ionic and electronic contributions,
\begin{equation}
e_{ij} = \frac{\partial P_i}{\partial \varepsilon_j} = e^{elc}_{ij} + e^{ion}_{ij}
\end{equation}
\begin{equation}
d_{ij} = \frac{\partial P_i}{\partial \sigma_j} = d^{elc}_{ij} + d^{ion}_{ij}
\end{equation}
where P$_i$, $\varepsilon_j$, and $\sigma_j$ denote the polarization vectors, strains, and stresses, respectively. The e$_{ik}$ is related to d$_{ik}$ by elastic tensor C$_{jk}$:
\begin{equation}
e_{ik} = d_{ij}C_{jk}
\end{equation}
By the Voigt notation, the Eq. (3) with C$_{4v}$ point group can be simplified as
\begin{equation}
\left(
\begin{array}{cccccccc}
0       &0       &0 \\
0       &0       &0 \\
e_{31}  &e_{31}  &0 \\
\end{array} \right) =
\left(
\begin{array}{cccccccc}
0       &0       &0 \\
0       &0       &0 \\
d_{31}  &d_{31}  &0 \\
\end{array} \right)
\left(
\begin{array}{cccccccc}
C_{11}  &C_{12}  &0      \\
C_{12}  &C_{11}  &0      \\
0       &0       &C_{66} \\
\end{array} \right)
\end{equation}
Hence, the out-of-plane piezoelectric coefficients d$_{31}$ can be obtained as,
\begin{equation}
d_{31} = \frac{e_{31}}{C_{11} + C_{12}}
\end{equation}
The calculated e$_{31}$ is 0.765 $\times$ 10$^{-10}$ C/m for the Se-Te interface bilayer V$_2$SeTeO, including the contributions of electrons 0.981 $\times$ 10$^{-10}$ C/m and the contributions of ions -0.216 $\times$ 10$^{-10}$ C/m. The d$_{31}$ is 0.57 pm/V obtained by formula Eq. (4). It's more than twice as large as that of monolayer V$_2$SeTeO \cite{38}. The large vertical piezoelectric polarization is anticipated to enable multifunctional piezoelectric devices based on the Se-Te interface bilayer V$_2$SeTeO.

\begin{figure}[htb]
\begin{center}
\includegraphics[angle=0,width=1.0\linewidth]{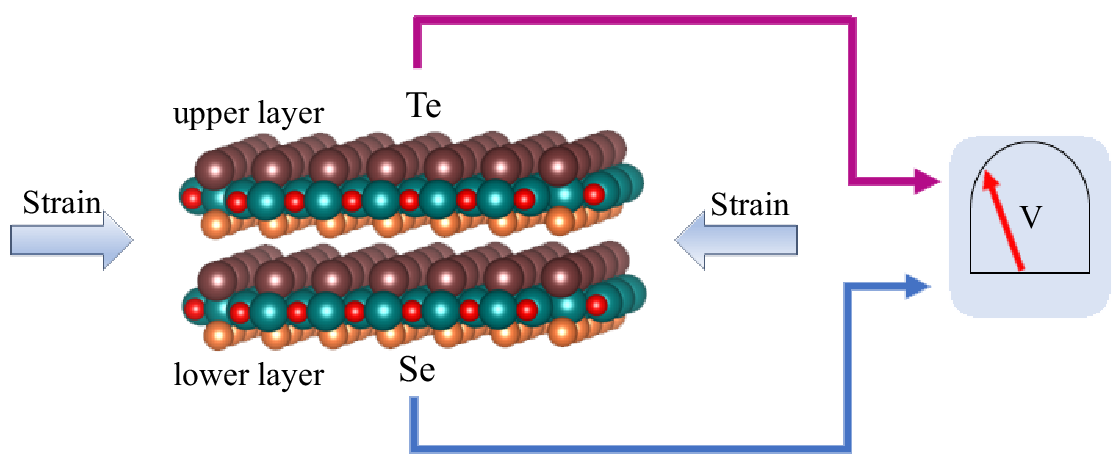}
\caption{ Schematic diagram of generating an out-of-plane piezoelectric effect in Se-Te interface bilayer V$_2$SeTeO.
}
\end{center}
\end{figure}

\section{CONCLUSION}
In conclusion, we propose a mechanism to manipulate the altermagnetic, multipiezo effect, and topological state in two-dimensional materials. Based on the analysis of symmetry and first principles calculations, the mechanism is verified in Janus bilayer V$_2$SeTeO. By transforming the stacking configuration, one not only can effectively tune the transform from AFM to altermagnetism, but also the phase transition from semiconductor to metal. More interestingly, the strain induces a novel multipiezo effect and topological state. The uniaxial strain can tune the magnitude of valley splitting, simultaneously, the valley polarization direction can be changed. The net magnetic moments generated by the piezomagnetic effect in the Se-Te interface bilayer V$_2$SeTeO. In addition, the Se-Te interface bilayer V$_2$SeTeO shows nodal loop under the 0$\%$ $\sim$ -3$\%$ uniaxial strain and the 0$\%$ $\sim$ -2$\%$ biaxial strain, while forms the Weyl points at near X and Y points under the further increased compressive strain. For the Se-Se/Te-Te interface bilayer V$_2$SeTeO, it becomes Dirac semimetal under the uniaxial/biaxial compressive strain of less than 2$\%$. Our work provides a platform to investigate the altermagnetic, multipiezo effect, and topological state in 2D altermagnets.

\section*{ACKNOWLEDGEMENTS}
This work is supported by the National Natural Science Foundation of China (Grants No. 12404076, No. 12474238, and No. 12004295). P. Li also acknowledge supports from the China's Postdoctoral Science Foundation funded project (Grant No. 2022M722547), the Fundamental Research Funds for the Central Universities (xxj03202205), and the Open Project of State Key Laboratory of Silicon and Advanced Semiconductor Materials (No. SKL2024-10), and the Open Project of State Key Laboratory of Surface Physics (No. KF2024$\_$02). W. Xun also acknowledge supports the Huai'an City Science and Technology Program Project (No. HAB2024067).


\end{document}